\documentclass[12pt,preprint]{aastex}
\begin{document}

\title{{\it XMM-Newton} and Optical Observations of Cataclysmic Variables from SDSS}

\author{Eric J. Hilton, Paula Szkody, Anjum Mukadam}
\affil{Astronomy Department, Box 351580, University of Washington, Seattle WA 98115}
\email{hilton,szkody,anjum@astro.washington.edu}

\author{Arne Henden, William Dillon}
\affil{American Association of Variable Star Observers, 49 Bay State Rd., Cambridge, MA 02138}

\author{Gary D. Schmidt}
\affil{Steward Observatory, University of Arizona, Tucson, AZ 85721}

\begin{abstract}
We report on {\it XMM-Newton} and optical results for 6 cataclysmic variables
that were selected from Sloan Digital Sky Survey spectra because they
 showed
strong HeII emission lines, indicative of being candidates for containing white
dwarfs with strong magnetic fields. While high X-ray background rates
prevented optimum results, we are able to confirm SDSSJ233325.92+152222.1
as an intermediate polar from its strong pulse signature at 21 min
and its obscured hard X-ray spectrum. Ground-based circular polarization
and photometric observations
were also able to confirm SDSSJ142256.31-022108.1 as a polar with a period near
4 hr.
Photometry of SDSSJ083751.00+383012.5 and SDSSJ093214.82+495054.7
solidifies the orbital
period of the former as 3.18 hrs and confirms the latter as
a high inclination system with deep eclipses. 
\end{abstract}


\section{Introduction}

While the Sloan Digital Sky Survey (SDSS; York et al. 2000) reveals new
cataclysmic variables (CVs) from its spectral database (Szkody et al. 2007 and
prior references), the correct identification of the type of close binary
often requires follow-up observations utilizing time series and 
 multiple wavelengths.
Optical photometry, polarimetry, and spectroscopy can identify the orbital
period, the spin period of the white dwarf and whether the object has
a magnetic field. The X-ray light curve, flux and spectrum can distinguish
a system with an accretion disk (dwarf nova or novalike) from those
systems containing a white dwarf with a strong magnetic field (polar, which
has the white dwarf spin synchronized to the orbit) or a lesser magnetic
field (intermediate polar (IP) with a white dwarf spin shorter than the orbital
period). For these magnetic systems in particular, the funneling of the
accretion to the magnetic pole or poles creates a strong modulation of the
X-ray fluxes (see Warner 1995 for a thorough review of the different types
of CVs and their multi-wavelength light curves). 

The high sensitivity and wide energy bandpass of {\it XMM-Newton}
have shown this satellite to be ideal for measuring the
X-ray fluxes and characterizing the X-ray source in the new 
CVs found in SDSS. The spin pulse and spectral temperature can be
identified in systems with moderate accretion rates, while flares and
softer spectral temperatures can be found in systems with extremely
low accretion rates (Szkody et al. 2004b, Schmidt et al. 2005, Homer et al.
2005, 2006). In 2005-2006, we were granted {\it XMM-Newton} time to observe
6 CVs with strong HeII emission lines that were recently found from
 SDSS spectra as
 good candidates for containing magnetic white dwarfs.
Unfortunately, except for SDSSJ233325.92+152222.2, 
the majority of the observations were conducted during times
of high X-ray background, with the result that 
only portions of the total times on the sources could be used and only
upper limits on the count rates could be obtained for 
 the other 5 systems SDSS J083751.00+383012.5,
SDSS J093214.82+495054.7, SDSSJ 154104.67+360252.9, SDSSJ 142256.31-022108.1 
and SDSS J204827.91+005008.9. 
 A request for 
re-observation
was denied. We include here the results for all systems,
 together with some
ground-based observations conducted along with X-ray observations 
that help elucidate
the nature of several of the objects. For the rest of the paper, we will use
abbreviations for the object names as SDSSJhhmm (hours and minutes of
right ascension). Table 1 provides a summary of the properties of the
6 systems that were under study. 

\section{Observations and Data Reduction}
\subsection{X-ray}

The two MOS detectors (Turner et al. 2001) 
and the EPIC pn detector (Stri\"uder et al. 2001) on {\it XMM-Newton} 
were used to obtain X-ray observations of the 6 cataclysmic variables.
In addition, the Optical Monitor (OM, Mason et al. 2001) obtained 
simultaneous optical observations, using the B filter (centered around 4500\AA).. The pipeline
products were used to create light curves and determine B magnitudes.
The data from the Reflection Grating Spectrograph were not useful due to
 low count rates.
The details of the X-ray and OM observations, including dates, UT times, observation
times (total time (TOT) and good time intervals (GTI) with background flaring
times removed),
and average count rates are listed in Table 2.

The data were reduced using the Science Analysis System (SAS) 
\footnote{http://xmm.esac.esa.int/sas/},
 ver. 7.0.0 with calibration files current to August 15th, 2006.
The data reduction followed the standard guidelines from the 
\emph{XMM-Newton} Web site (VILSPA) and from the NASA/GSFC \emph{XMM-Newton} 
Guest Observer Facility ABC Guide (ver. 2.01). The total observation data were
used to identify the source visibility and background levels and then the
data were screened to eliminate the high background where possible in order
to increase the signal to noise.
New event list files were created directly from the observation data files using the SAS tools.
From these event list files, light curves were created in the 
10-18 keV range using the entire chip for each detector and these files were 
used to identify and remove background flaring events.
Each of the targets 
had different criteria for selecting good data intervals, depending on their 
individual light curves.
For SDSSJ2333, SDSSJ2048 and SDSSJ1541, the high background count rates were most 
easily identified by using a count rate cut-off.
For SDSSJ2333, data were ignored when the count rate exceeded 3.0 
counts s$^{-1}$ for the pn and 0.6 counts s$^{-1}$ for both MOS detectors.
For SDSSJ2048, we used limiting values of 12.0 counts s$^{-1}$ for the pn 
and 
5.0 counts s$^{-1}$ for both MOS detectors as
the background 
counts were higher throughout the observation. The limit used for
SDSSJ1541 was 5 counts s$^{-1}$ for both pn and MOS detectors.
For the other three targets, SDSSJ0837, SDSSJ0932 and SDSSJ1422,
 the background rate 
was stable throughout most of the observation, but increased
significantly at either the beginning or the end of the observation.
It was therefore easier to trim the data so that only data during the
stable background times was used.

Five of our observations (SDSSJ0837, SDSSJ0932, SDSSJ1422, SDSSJ1541 and 
SDSSJ2048) had no obvious visible source.
We note that in the case of
SDSSJ1541, several bright pixels appear in the pn image somewhat near the 
expected location of the
source. This is not apparent in the MOS, even though background sources that
are fainter in the pn image than these bright pixels do appear in the MOS
images. Additionally, the bright pixels do not show the psf of other
sources visible in both detectors so we concluded that this was not the target.
 For these objects, we
employed the following method to place an upper limit on the count rate. Using
the pn detector images because of their increased sensitivity over the MOS, we used
the target RA and Dec to place an aperture (360 pixel radius), 
and then found the number of counts 
inside
this target aperture, using an energy range of 0.1-15 keV. We then
 found the number of counts inside four
circular background apertures of the same size, 
 one in each quadrant. Each background aperture was
a large area free of point sources located at a similar distance from the 
midplane of
the detector.  
For our observations, the exposure time on the 12 pn chips
varied slightly, so a count rate was determined for each aperture. We then
calculated the count rate needed to have a 3$\sigma$ detection, added
this to the background, and compared to the count rate for the target
aperture. In other words, given a measured background, we calculated
the count rate needed to reach a given signal-to-noise, using:

\begin{equation}
\sigma= \frac{n\sqrt{t}}{\sqrt{n+b}}
\end{equation}

where $\sigma$ is the signal-to-noise, $n$ is the count rate needed to reach
that signal-to-noise, $t$ is the time, and $b$ is the background count rate.
We used the average of the count rate needed for a 3$\sigma$ detection in the
four background apertures as our limit.
As a consistency check, this analysis was carried out on a field source that
appeared by eye to be very faint.
The field source was a 5$\sigma$ detection.

In order to ensure that real signal had not been removed when the flaring
background sections were discarded, we also repeated this analysis keeping
all of the exposure and limiting the energy range to 0.5-2.0 keV. In this
analysis, no target was visible by eye in the
image, and no target was above 3$\sigma$. 

For the source that was successfully detected, SDSSJ2333, the event list files were filtered 
with the standard canned expressions, and were restricted to the well-calibrated
 regions of 0.2-15 keV for spectral analysis and 0.1-12.5 keV for light-curve 
analysis.
With the pn detector, only single events (pattern = 0) were accepted for 
spectral analysis, and singles and doubles (pattern $\leq$ 4) for light-curve 
analysis.
For the MOS detectors, quadruples and lower (pattern $\leq$ 12) were permitted 
for both the spectral analysis and the light-curve analysis.
The data were binned to 20 counts per bin to increase
 the signal-to-noise ratio for the spectral analysis.
The source region was taken to be a circular aperture with radius of
 360 pixels for 
both detectors.
The background region for the MOS detectors was determined from an annulus 
centered on the target and free of other 
sources while the pn background was determined using four rectangular regions,
each on adjacent CCD chips, with similar Y locations as the target.

The resulting lightcurve is background-subtracted, and the times have been corrected to 
the solar system barycenter using FTOOLS (Blackburn 1995) 
\footnote{tools available at http://heasarc.gsfc.nasa.gov/lheasoft/ftools/}.
The pn lightcurve data as well as the OM data were binned at 
200 seconds to increase the signal-to-noise ratio.

SDSSJ0932 was the only other source to have sufficient counts in the OM to 
show variability on the orbital period. For this object, the OM data were
binned at 60s to increase the S/N.

\subsection{Ground-based data}

Optical observations on SDSSJ0837, SDSSJ0932, SDSSJ1422 and SDSSJ2048
 were conducted at the US Naval
Observatory Flagstaff Station (NOFS), as well as at the George Observatory. The
optical photometry summary can be found in Table 3. 

For SDSSJ0932, the NOFS observations were made with the automated 1.3m
wide-field telescope and a single 2048x4096 E2V back-illuminated CCD
without any filter.  This combination gives a pixel scale of 0.60arcsec/pix
and a field of view of 20x40 arcmin.
The exposure times were 90 seconds, with 50 seconds
of dead-time between exposure.  For the other NOFS targets, the 1.0m
telescope and a SITe/Tektronix 102x1024 back-illuminated CCD without
filter were used.  This combination gives a pixel scale of 0.67arcsec/pix
and a field of view of 11.3x11.3 arcmin. Exposure times averaging
200 seconds were used; deadtime at the 1.0m was always 40 seconds.
In all cases, images were processed using standard
techniques, and aperture photometry was performed using DAOPHOT
(Stetson 1987) as implemented in IRAF.  The exception was for SDSSJ2048,
where the conditions were poor and DAOPHOT psf-fitting was used to
obtain the highest signal/noise.

Once instrumental magnitudes were obtained, inhomogeneous ensemble
photometry
techniques following the guidelines of Honeycutt (1992) were used to obtain
the
final photometry.  At least 10 stars in each field were used for the
ensemble.  Each error estimate is the quadrature sum of the
Poisson and the ensemble errors.

All four optical targets have SDSS calibration photometry, but we also
obtained B and V optical calibration at NOFS on photometric nights,
using standards from Landolt (1983, 1992).  The V-band calibration
was used for the photometry shown in the figures and analysis, as the
CCD response was closest to this passband.  The calibration photometry
is available on the AAVSO website\footnote{http://www.aavso.org/}
 and through their on-line comparison
star database.

The George Observatory measurements were made with a 0.46cm telescope and
SBIG CCD in cirrus sky conditions. The images were stacked in 5 minute bins
and magnitudes were measured in comparison to other stars in the field to
 determine a differential light curve. 

Circular polarization was also measured for SDSSJ1422 using the
CCD Spectropolarimeter SPOL on the 2.3m Bok Telescope on Kitt Peak on
the nights of 2005 March 16 and April 14. 
The instrument configuration and data acquisition was as
described in Schmidt et al. (1992), 
with the exceptions of an improved quality camera
lens and state-of-the-art 1200$\times$800 pixel cosmetically perfect SiTE 
CCD with 2.2 e$^-$
read noise that was thinned, back-illuminated, and anti-reflection coated by 
the University of
Arizona's Imaging Technology Laboratory.

\section{Results}

\subsection{SDSSJ2333}

This object was discovered
and identified as a likely intermediate polar (Szkody et al. 2005; Sz05) 
from 9 hrs of 
photometry in 2004 that showed
a 21 min period high amplitude modulation in its light curve,
while an orbital period of 1.4 hrs was determined from a radial
velocity curve obtained from 3.5 hrs of
spectra. Southworth et al. (2007; Sw07) recently reported 4 hrs of 
photometry  
and spectroscopy covering 6 hrs over 2 nights in 2006. While their spectra
confirmed and refined the period to 83.12 min, the photometry was 
consistent with a double-sinusoid variation indicating a spin period of
41.66 min and its harmonic at 21 min. This result
 implies two accreting poles causing the variation.
The confirmation of an IP generally consists of a large amplitude variation
at the spin period in X-ray, combined with a hard, absorbed X-ray spectrum.
Fortunately, the best {\it XMM-Newton} data among the 6 objects were obtained for this
source.

Figure 1 shows the light curve in both the optical OM (top panel) and X-ray pn (bottom panel).
A prominent modulation is obvious in both light curves.
The discrete Fourier transform (DFT) of these data, shown in Figure 2, reveals
 a period of 21 minutes (760 $\mu$Hz), consistent with that found in the 
previous optical observations by Sz05. The period of 41.66
min (400 $\mu$Hz) found by Sw07 is not evident. It
is not clear if these differences are due to 
changes in the two accreting 
poles. The Sz05 optical dataset was obtained with a 1m telescope,
using 140s integrations but was 9 hrs long (encompassing about 13 cycles if
the period is 42 min). The Sw07 dataset was obtained with
a larger (2.2m) telescope and 40s integrations but was only about 4 hrs long
(somewhat less than 6 cycles of 42 min). Their unfolded light curve shows
large variations in the depths of the minima so a few cycles could
influence the resulting pattern. Our XMM data are even shorter (about 3 hrs
or only 4 cycles of 42 min) and the peaks and troughs are highly variable.
Figure 3 shows the {\it XMM-Newton} data folded on the 41.66 min
period. Due to the scatter, any difference in the two peaks in each cycle
are not obvious.

We did fold the Sz05 optical dataset on 41.66 min and
the resulting light curve shows two similar humps per cycle (Figure 4), unlike 
the 
light curve shown
 in Figure 4 of Sw07.
The DFT of the Sz05 data (Figure 5) shows the orbital period (0.0002 Hz)
and the prominent 21 min period but only a small insignificant peak that would be
consistent with the 41.66 min period,
in contrast to the Sw07 data which have a narrower and higher
amplitude peak for the 42 min than the 21 min period.  While the 
{\it XMM-Newton} data only show a 21 min period, the
larger X-ray amplitude (2-3 times larger than the optical) confirms the
IP nature of this source. Compared to the XMM results on the IP SDSS J1446+02,
where the X-ray spin pulse maximum occurs 0.3 phase before the optical (Homer et al. 2006), the X-ray and optical pulses in SDSS J2333 appear to
be in phase.

As further confirmation of the IP nature, we folded the data on the 41.66 min
period using different
energies (Figure 6 shows the resulting light curves). 
  As is typical for IPs, 
the amplitude of the variability increases for lower energies, likely due to
a local absorber in the system.

The pn spectrum of SDSSJ2333 is shown in Figures 7 and 8, binned at 20 counts per 
bin to increase the signal-to-noise ratio.
The best fit XSPEC\footnote{available at http://heasarc.gsfc.nasa.gov/docs/xanadu/xspec/index.html} simple
absorbed bremsstrahlung model gave a 
reduced $\chi^2$ = 1.07, with a hydrogen column density of the
 absorbing gas of 9.2$\times$10$^{20}$ cm$^{-2}$ and a temperature of 5.1$\pm$1.4 keV (Table 4). 
While this simple model provided a reasonable fit with a low reduced 
$\chi^2$, 
there was excess emission near
1 keV that could be due to a complex of Fe, Ne, O emission lines that is present in
IPs (Mukai et al. 2003) and unresolved in our data. However, the fit with a mekal
model (which includes emission lines) produced similar values (see Table 4) of
temperature and column but does not fit the excess (Figure 7) either.
While the spectrum is typical
of IPs in showing a large absorption, the spectrum is not as hard as
usual (10-20 keV), although it does not show a
 prominent soft X-ray component as evident
in increasing numbers of IPs (Haberl et al. 2002; deMartino et al. 2006).
We next tried partial covering absorbers (pcf) and cooling flow models
(mkcflow). There was little improvement in the pcf model but the cooling
flow seemed to fit the spectrum best (Table 4 and Figure 8). This 
type of model was used by Mukai et al. (2003) to fit the IP EX Hya.

\subsection{SDSS J0837}

Sz05 identified SDSSJ0837 as a polar candidate showing a cyclotron hump
in the blue and TiO bands from the secondary star in the red. Follow-up
polarimetery and spectroscopy (Schmidt et al. 2005) confirmed the polar
nature, determined a spectral type of M5V, a distance of 330 pc and an
orbital period of either 3.18 or 3.65 hr.
Analysis of the spectrum showed this object has a very low accretion rate,
similar to systems like MQ Dra (SDSSJ1553+5516)
 and SDSSJ1324+0320 (Szkody et al. 2003).  {\it XMM-Newton}
observations of these latter 2 systems (Szkody et al. 2004a) 
showed very low X-ray fluxes (10$^{-14}$ ergs cm$^{-2}$ s$^{-1}$)
and luminosities ($<$10$^{29}$ ergs s$^{-1}$),
and soft spectra (0.2-0.8 keV) 
characteristic of the secondary M star rather than a shock
from accretion onto the white dwarf. While the high background rates 
combined with low source counts prevented
a detection of SDSSJ0837 in our observations, the upper limit gives us a
clue as to the nature of the X-ray flux. Table 5 compares the
secondary spectral types, distance and pn count rates for the 3 sources.
If the X-ray spectrum is similar to MQ Dra (kT=0.8 keV), then scaling by the
distance squared predicts a count rate of 0.003 for SDSSJ0837, which is
below the limit from our observation.  The limit of 0.0042 for SDSSJ0837
would correspond to a 0.2-10 keV flux of $<$8$\times$ 
10$^{-15}$ ergs cm$^{-2}$ s$^{-1}$ and an X-ray luminosity of $<$10$^{29}$ ergs s$^{-1}$.Thus, it appears that there is
also no active X-ray shock in SDSSJ0837 and this system is indeed in the
regime of extremely low accretion rate.

The count rate from the OM B filter gives a magnitude of 19.4, at the time
of the X-ray observation, but there is no obvious
variability in the OM light curve.
Our optical photometry spans times a year prior to the X-ray data and 3
months after (Tables 2 and 3). Both data sets show the object at the same
faint magnitude as evident in the SDSS photometry and that of Schmidt et
al. (2005). Our light curves (Figure 9) show a repetitive structure (typical of
polars) but with a low amplitude of modulation (as noted by Schmidt et al.
and suggested to be due to a low inclination of the system). The spacing
of the lowest points of the light curve are 3.14$\pm$0.10 and 3.00$\pm$ 0.10
hours for the 2 nights; Figure 9. Running a discrete Fourier transform on
the data reveals periods of 2.98$\pm$0.08 for Dec 15 and 3.06$\pm$0.06 for
Jan 27. Both of these methods show
that the shorter period is
the correct orbital period for this system.

\subsection{SDSSJ0932}

The SDSS spectrum of SDSSJ0932 is highly unusual (Szkody et al. 2006) with
a very strong, asymmetric HeII 4686 emission line superposed on a blue continuum
with only weak Balmer emission. Little information on this object exists
in the literature, other than a discovery spectrum by Munari et al. (1997)
showing stronger Balmer emission and an abstract report (Holcomb et al. 1994)
reporting weak ROSAT sources. The Center for Backyard Astrophysics
 web site contains a light curve made
by Robert Fried\footnote{http://cbastro.org/results/highlights/uma6/; note that
this is a differential light curve that may not be calibrated with known
standard stars} that
shows an eclipse and
gives a period of 10.04 hrs, which is very long for a cataclysmic variable.
Our {\it XMM-Newton} observation provided an upper limit of 0.014 c/s,
confirming a low X-ray flux from this object, despite the strong HeII presence
and its relatively bright optical mag (g$\sim$17.5). 

The OM data in the B filter show that an eclipse took place during the latter
half of the pn observations (Figure 10). We also tried filtering the times
of the pn data to exclude the eclipse, but still no source was visible.
Our optical photometry obtained about 2 months prior to the X-ray observations
is shown in Figure 11. A deep optical eclipse is evident which is
offset to earlier phases from the center of an overlying hump. This offset appears
different than the Fried light curve (where the eclipse appears centered in
the hump structure), indicating a moving structure of
the underlying emission region.
The eclipse depth in B appears to be about 0.5 mag deeper than in V,
as is typical for eclipses of a hot white dwarf and inner disk (Baptista,
Steiner \& Cieslinski 1994).

\subsection{SDSSJ1422}

This object was first identified in the Two-Degree Field (2dF) QSO survey
(Croom et al. 2001) and suggested to contain a magnetic white dwarf by
Marsh et al. (2002). The latter authors identified a possible 3.37 hr orbital
period (with possible 1 cycle/day aliases) from spectroscopic data. This object also exists in the SDSS database (Szkody et al. 2004a)
and shows similar spectra to those obtained by Marsh et al. The
high X-ray background on this source resulted in very limited data so
we were only able to place an upper limit on the X-ray flux. In
addition, the source was too faint for any detection in the OM.
However, our ground-based spectropolarimetry
(Figure 12) obtained on two different nights (2005 March 16 for 960 sec 
and April 14 for 3600 sec) show coadded circular polarization values of
+2.5\% and +2.8\% respectively, thus confirming this object as a polar.
Polarization is present throughout the duration of the longer observation,
which was obtained in three segments, and each segment displays a net
circular polarization betwenn +1.5\% and +4\%, with individual error bars
of $\sim$0.5\%. From the coadded results shown in Figure 10, it also can
be seen that the polarization is approximately uniform over the optical
spectrum, suggesting the smeared cyclotron harmonics of a high-temperature
shock (characteristic of a bonafide polar), as opposed to the isolated
emission features of a low-accretion rate system.
While the continuum flux level in 2005 looks similar
to the SDSS spectrum from 2002, there is a rise at the blue end and the
emission lines are not as strong, especially the HeII emission line. 
Thus, the accretion
state may have been slightly different than in the earlier data.

While our optical photometry was obtained several years prior to the
X-ray observations, we show the light curve
 in Figure 13 as there has not been any
prior photometry published that we could find. There is a sinusoidal
type modulation with full amplitude of $\sim$ 0.4 mag on each night
and pdm analysis gives a period of 4 hrs, but since this period
is about the length of the data string on each night and the nights are
too far apart to combine the data to obtain a better period, this value
must remain as tentative. However, the photometry indicates the period
could be longer than the estimate provided by Marsh et al. (2002).

\subsection{SDSSJ2048}

SDSSJ2048 is very similar to SDSSJ0837. Schmidt et al. (2005) provided
the discovery information, identifying it as another very low accretion
rate polar with an orbital period near 4.2 hr, an M3 secondary,
 and a distance of 260 pc. 
The {\it XMM-Newton} observations have a high background count over the whole chip, but the top half of the chip
has systematically higher counts than the bottom half of the chip.
Although we provide an upper limit, researchers should be cautioned that there 
may have been a problem with the pn detector during this observation.
As with SDSSJ0837, there was no significant detection of the source, leading to
a limit to the count rate of $<$0.013. Using the same arguments as for
SDSSJ0837 (Table 5), the expected count rate would be 0.005 c/s for a
comparable spectrum as for MQ Dra. This is also consistent with our limit
and argues for a lack of shock at the magnetic pole of the accreting white dwarf.
The XMM limit would translate into a 0.2-10 keV flux of 
$<$3$\times$10$^{-14}$ ergs cm$^{-2}$ s$^{-1}$ and an X-ray
luminosity of $<$2$\times$10$^{29}$ ergs s$^{-1}$.

The OM data show a very low count rate (Table 2) that translates into
a B magnitude of 19.9.
Our optical photometry obtained about 6 months prior to the X-ray observations
appears to be about one magnitude brighter than the SDSS photometry but the
observations were made in cirrus and the error bars are large. The light
curve (Figure 14) shows variability of 0.3 mag which may be related to the 
orbital cycle but better photometry is needed to confirm this.

\subsection{SDSSJ1541}

This object was identified as a polar in Sz05. While the SDSS spectrum was
obtained during a time of low accretion, the follow-up spectropolarimetry
and photometry were obtained during high accretion states. During a high
state, the X-ray flux is expected to be large and to show both soft and
hard components. The OM data show the object at a low state, B$\sim$20 mag.
Unfortunately, the high radiation background during the
{\it XMM-Newton} observation only left sufficient time to provide a limit
of 0.0160 c/s. Combined with the unknown distance and spectrum, this value
cannot constrain the source of emission.

\section{Conclusions}

Although high background rates prevented good observations of our
faintest targets, the {\it XMM-Newton} pn observations have been
able to confirm the IP nature of SDSSJ2333, and the limits on the fluxes
of the very low accretion rate polars SDSSJ0837 and SDSSJ2048 show
that these two systems have very low X-ray fluxes and luminosities ($<$10$^{29}$
ergs s$^{-1}$) indicative of a lack
of related 
accretion shocks. Our ground-based spectropolarimetry on SDSSJ1422
confirms it as a polar system and our OM and ground photometry on 
SDSSJ0932 provide
good eclipses for this long period novalike system. While the presence
of strong HeII emission was the initial defining characteristic of these 6 
systems, the followup observations at other wavelengths have shown a wide
variety of behavior.

\acknowledgements

This work was supported by \emph{XMM-Newton} grants NNG05GL29GG and
NNX06A38OG to the University
of Washington and is based on observations obtained with \emph{XMM-Newton},
an ESA science mission with instruments and contributions directly funded
by ESA Member States and the USA (NASA). We thank the members of the AAVSO
who contributed observations to determine the state of the systems close
to the X-ray observations, especially Gary Walker. We also thank S. Levine
for his help in making the NOFS 1.3m observations, and to the USNO for the
use of their facilities.

\clearpage
\begin{deluxetable}{lcclcl}
\tablewidth{0pt}
\tablecaption{Summary of Objects}
\tablehead{
\colhead{Object} & \colhead{g mag} & \colhead{P$_{orb}$ (hr)} & \colhead{Type}
& \colhead{d(pc)} & \colhead{Ref\tablenotemark{a}}}
\startdata
SDSSJ0837 & 19.13 & 3.18 & Polar & 330 &  Sz05,Sch05 \\
SDSSJ0932 & 17.45 & 10.0 & NL & ... & S06 \\
SDSSJ1422 & 19.84 & 3.37\tablenotemark{b} & Polar & ... & M02,S04 \\
SDSSJ1541 & 19.70 & 1.4 & Polar & ... & S05 \\
SDSSJ2048 & 19.38 & 4.2 & Polar & 260 & Sch05,S06 \\
SDSSJ2333 & 18.75 & 1.4 & IP & ... & S05 \\ 
\enddata
\tablenotetext{a}{M02= Marsh et al. 2002; S04=Szkody et al. 2004; Sz05=Szkody et al. 2005;
Sch05=Schmidt et al. 2005; S06=Szkody et al. 2006}
\tablenotetext{b}{M02 state this period could have 1 cycle/day aliases}
\end{deluxetable}

\clearpage
\begin{deluxetable}{llllrllll} 
 \tabletypesize{\scriptsize}
\tablewidth{0pt}
\tablecaption{X-ray Observations}
\tablehead{
\colhead{Target} & \colhead{Date} & \colhead{Instr} & \colhead{Filter} &
\colhead{Time\tablenotemark{a} (s)} &\colhead{UT Start} &\colhead{UT Stop}
& \colhead{Ave Count rate\tablenotemark{b} (C/s)} & \colhead{3$\sigma$ upper limit}}
\startdata
SDSSJ2333 & 2006 Jun 04 & pn   & Thin1 & 10737/10436 & 09:00:47  &  11:59:44 &  0.184$\pm$.005 & \\
&&MOS1 & Thin1 & 12372/10866 & 08:38:28  &  12:04:40 & 0.078$\pm$0.003  &\\
&&MOS2 & Thin1 & 12379/11271 & 08:38:28  &  12:04:47 & 0.074$\pm$0.003 & \\
&&OM   & B     & 7200  & 09:55:16  &  12:03:42 & 1.3$\pm$0.4 (B=19.0) & \\

SDSSJ2048 & 2005 Apr 21 & pn   & Thin1 & 15829/7570 & 14:25:45  &  18:49:34 & &  0.0138  \\
&&OM & B & 16080 &   &  & 0.5$\pm$0.4 (B=19.9) & \\
SDSSJ0837 & 2005 Oct 07 & pn   & Thin1 & 12439/9700 & 09:38:52  &  13:06:11 & &  0.0042  \\
&&OM & B & 12961 & & & 0.8$\pm$0.3 (B=19.4) & \\
SDSSJ0932  & 2006 May 16 & pn   & Thin1 &  9581/8385 & 19:01:11  &  21:40:52 & &  0.0148  \\
&&OM & B & 10196 & & & 10.5$\pm$1.5 (B=16.71)\tablenotemark{c} & \\
SDSSJ1541 & 2005 Aug 24  & pn  & Thin1 & 9628/4855 & 09:17:59 & 12:46:19 & & 0.0160 \\
&&OM & B & 10179 & & & 0.4$\pm$0.1 (B=20.0) & \\
SDSSJ1422 & 2006 Feb 07 & pn & Thin1 & 5702/1996 & 21:08:43 & 23:33:03 & & 0.0128 \\
\enddata
\tablenotetext{a}{pn,MOS times are Total time/Good Time Interval, OM times are from pipeline products}
\tablenotetext{b}{X-ray count rates determined from spectral reductions;
OM count rates and B magnitude from pipeline products with 
error bar representing the variance}
\tablenotetext{c}{magnitude is outside of eclipse}
\end{deluxetable}

\clearpage
\begin{deluxetable}{lccc}
\tablewidth{0pt}
\tablecaption{Optical Photometry}
\tablehead{
\colhead{Object} & \colhead{Obs} & \colhead{UT Date} & \colhead{Time} }
\startdata
SDSSJ0837 & USNO & 2004 Dec 15 & 07:49-13:16 \\
SDSSJ0837 & USNO & 2006 Jan 27 & 06:47-12:05 \\
SDSSJ0932 & USNO & 2006 Feb 27 & 02:32-10:03 \\
SDSSJ0932 & George & 2006 May 12 & 02:48-04:27 \\
SDSSJ1422 & USNO & 2003 Jun 27 & 03:48-07:42 \\
SDSSJ1422 & USNO & 2003 Jun 30 & 03:44-07:28 \\
SDSSJ2048 & USNO & 2004 Sep 05 & 02:52-09:13 \\
\enddata
\end{deluxetable}

\clearpage
\begin{deluxetable}{lcccc}
\tablewidth{0pt}
\tablecaption{Model Fits to SDSSJ2333}
\tablehead{
\colhead{Model} & \colhead{Reduced $\chi^2$} & \colhead{N$_{H}\times$10$^{20}$ cm$^{-2}$} & \colhead{kT (keV)} & \colhead{Norm} }
\startdata
bremss & 1.07 & 9.2$\pm$1.6 & 5.1$\pm$1.4 & 2.2$\times$10$^{-4}$ \\
mekal & 1.03 & 8.1$\pm$1.6 & 5.2$\pm$1.0 & 5.5$\times$10$^{-4}$ \\
pcf+mekal\tablenotemark{a} & 1.05 & 8.1$\pm$1.6 & 4.9$\pm$1.4 & 6.2$\times$10$^{-4}$ \\
mkcflow & 0.81 & 8.8$\pm$1.6 & 0.4-17\tablenotemark{b} & 7.9$\times$10$^{-16}$ \\
\enddata
\tablenotetext{a}{pcfabs fixed at 2$\times$10$^{23}$; resulting covering fraction of 13\%}
\tablenotetext{b}{low and high temperatures of cooling flow}
\end{deluxetable}

\clearpage
\begin{deluxetable}{lccl}
\tablewidth{0pt}
\tablecaption{Comparison of Low Accretion Rate Objects}
\tablehead{
\colhead{Object} & \colhead{Sec} & \colhead{d (pc)} & \colhead{pn C/s} }
\startdata
MQ Dra & M5 & 130 & 0.020 \\
SDSSJ1324 & M6 & 450 & 0.0012 \\
SDSSJ0837 & M5 & 330 & $<$0.0042 \\
SDSSJ2048 & M3 & 260 & $<$0.0138 \\
\enddata
\end{deluxetable}

\clearpage
\begin{figure}
  \epsscale{.80}
  \plotone{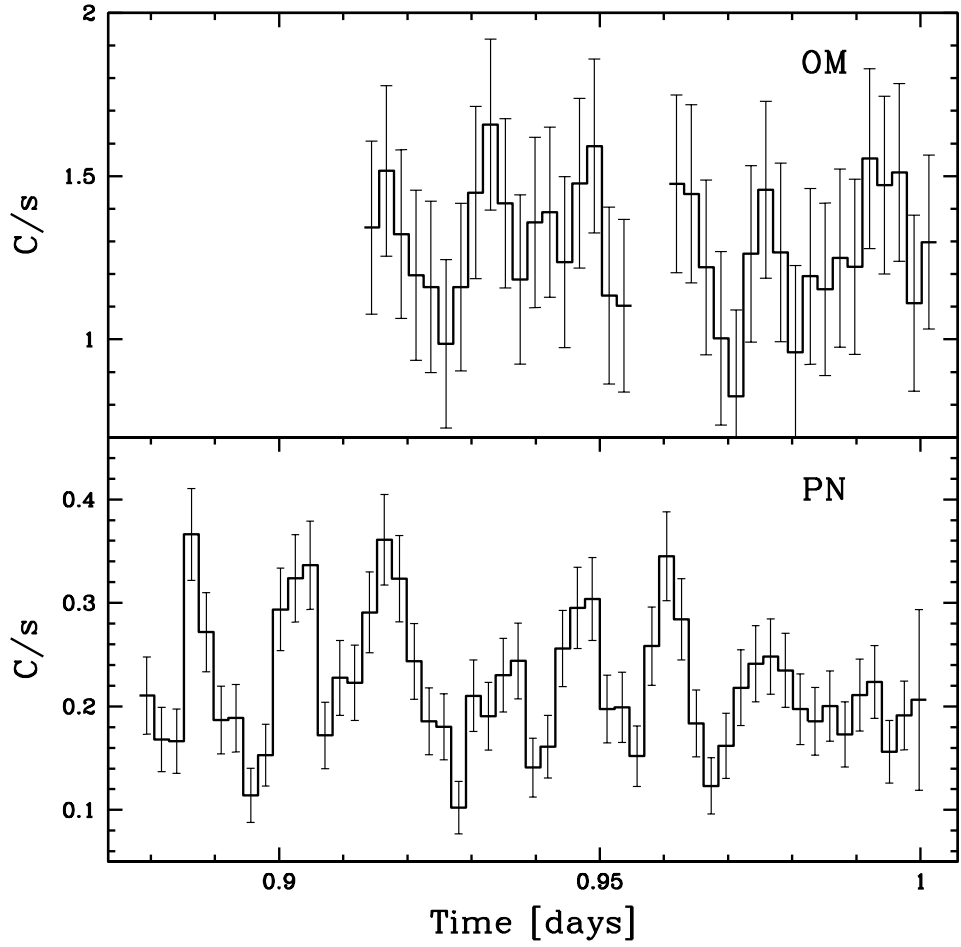}
  \caption{OM and pn light curves of SDSSJ2333 showing the strong modulation
in both optical and especially X-ray.}	
\end{figure}

\clearpage
\begin{figure}
  \epsscale{.80}
  \plotone{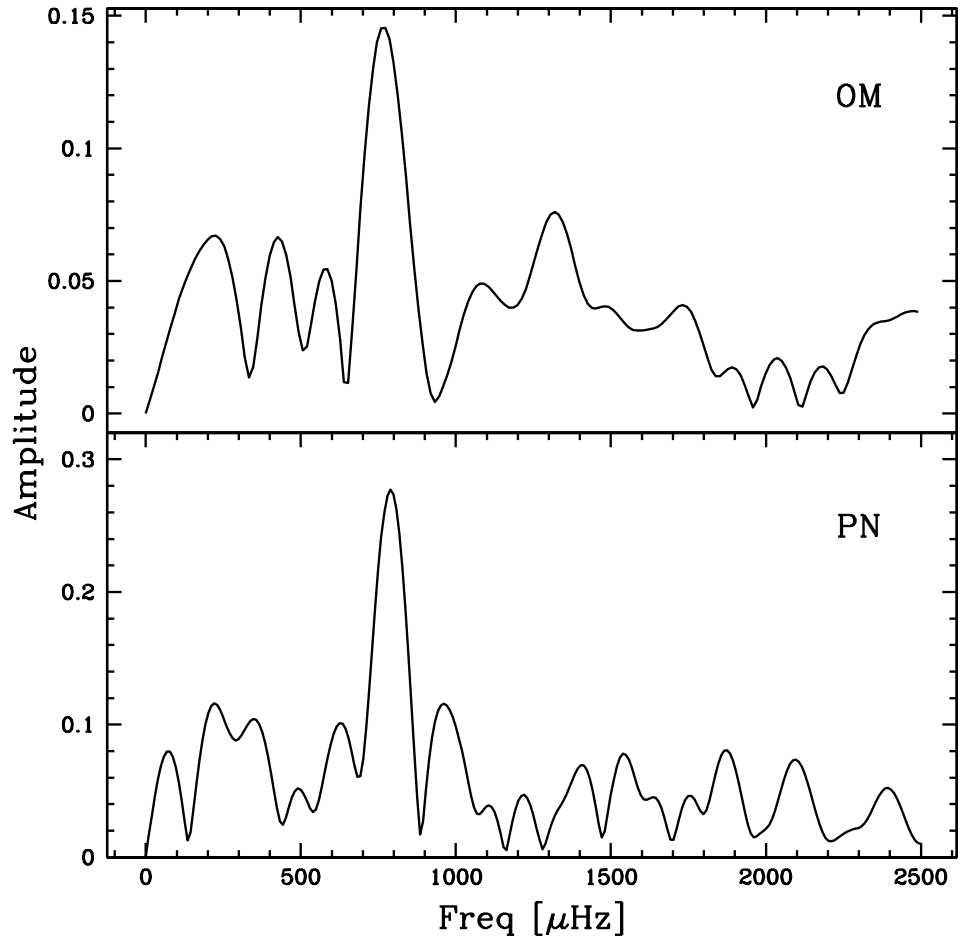}
  \caption{DFT of SDSSJ2333 showing the prominent period at 21 min and its 
large amplitude of modulation. Note there is no peak evident at the
41.66 min period (400$\mu$Hz) found by Southworth et al. (2007).}
\end{figure}

\clearpage
\begin{figure}
\plotone{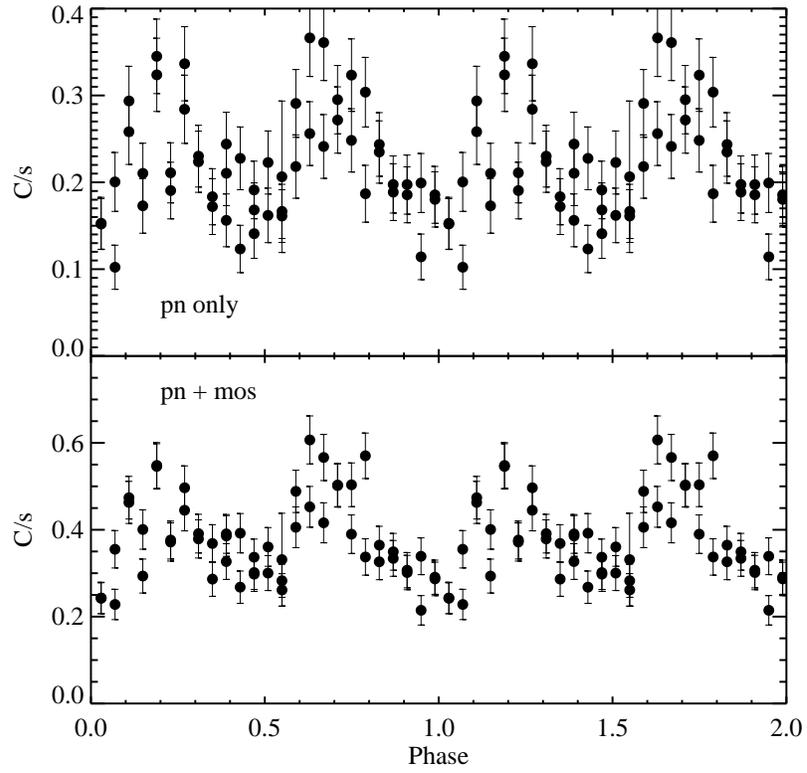}
\caption{{\it XMM-Newton} pn and pn+mos data on SDSSJ2333 folded on the 41.66 min period.}
\end{figure}

\clearpage
\begin{figure}
\plotone{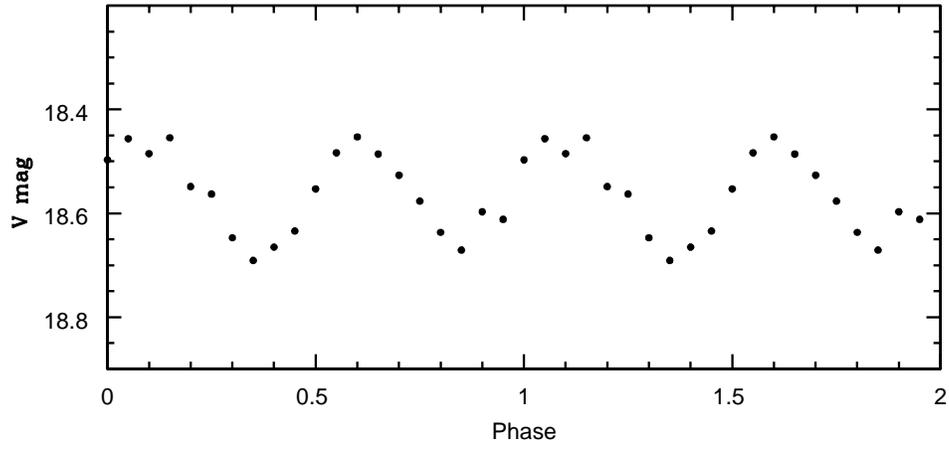}
\caption{Optical data of SDSSJ2333 from Sz05 phased on the 41.66 min spin period.}
\end{figure}

\clearpage
\begin{figure}
\epsscale{0.80}
\plotone{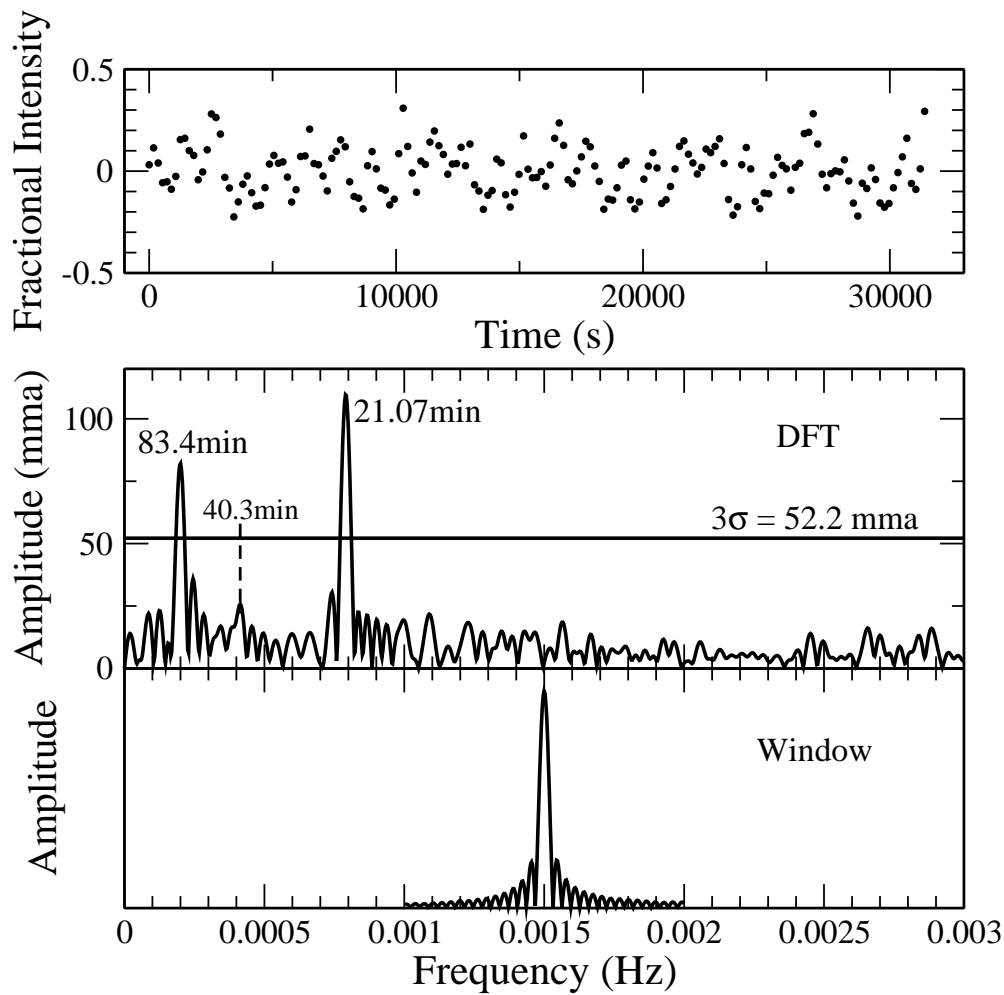}
\caption{DFT of the optical data from Sz05. The 21 min
period is evident but the 42 min period is not significantly detected above the
noise. Amplitude units are milli modulation amplitudes. Top panel is the light
curve divided by its mean (in intensity units) minus 1. Bottom panel is the
transform of a noiseless frequency with the same time sampling as the data in order
to show the alias pattern due to the finite length and gaps in the data.}
\end{figure}

\clearpage
\begin{figure}
  \epsscale{.80}
    \plotone{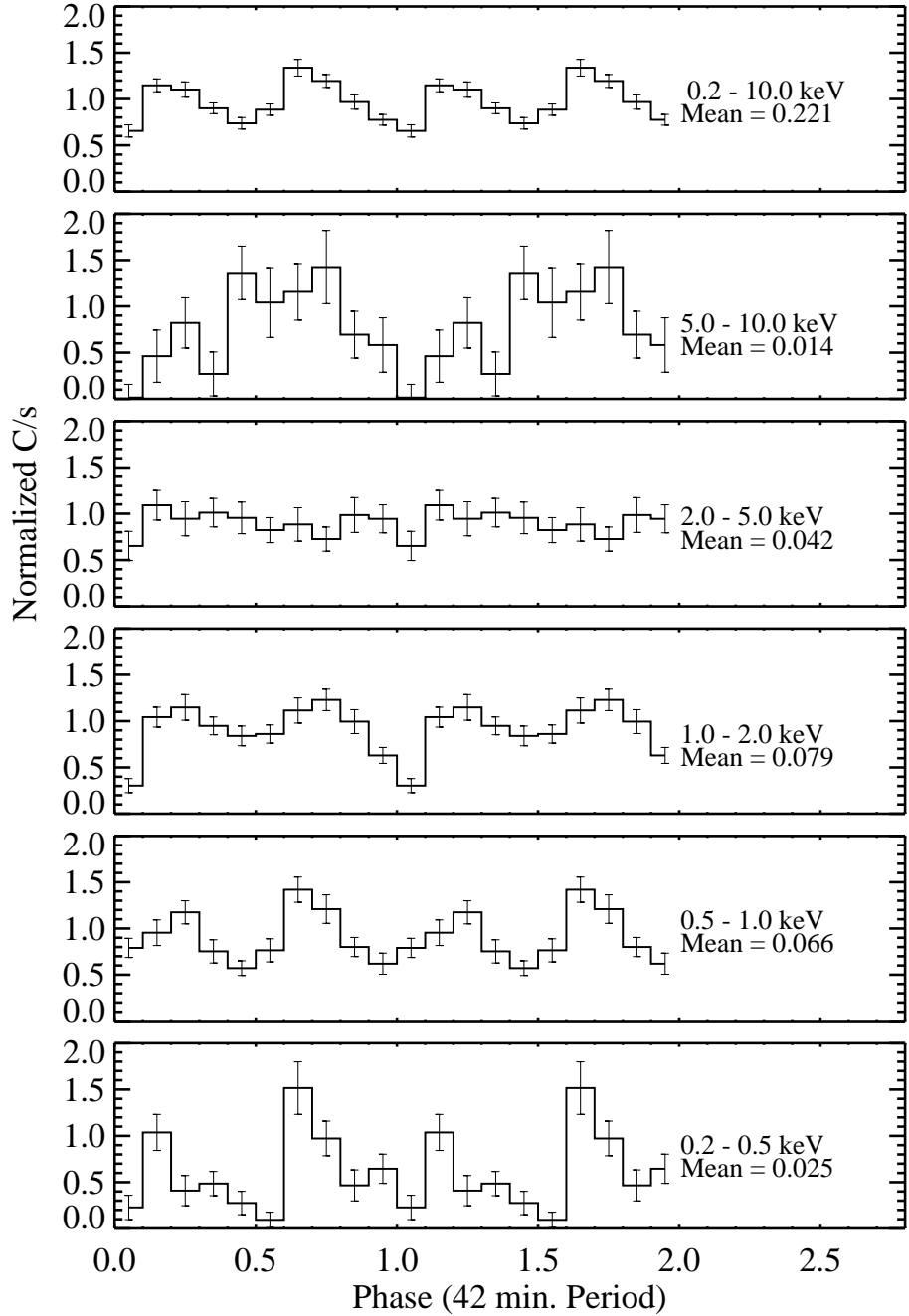}
  \caption{PN data of SDSSJ2333 at different energies and phased on the 41.66 min period.}
\end{figure}

\clearpage
\begin{figure}
  \epsscale{.80}
  \includegraphics[angle=270,scale=.60]{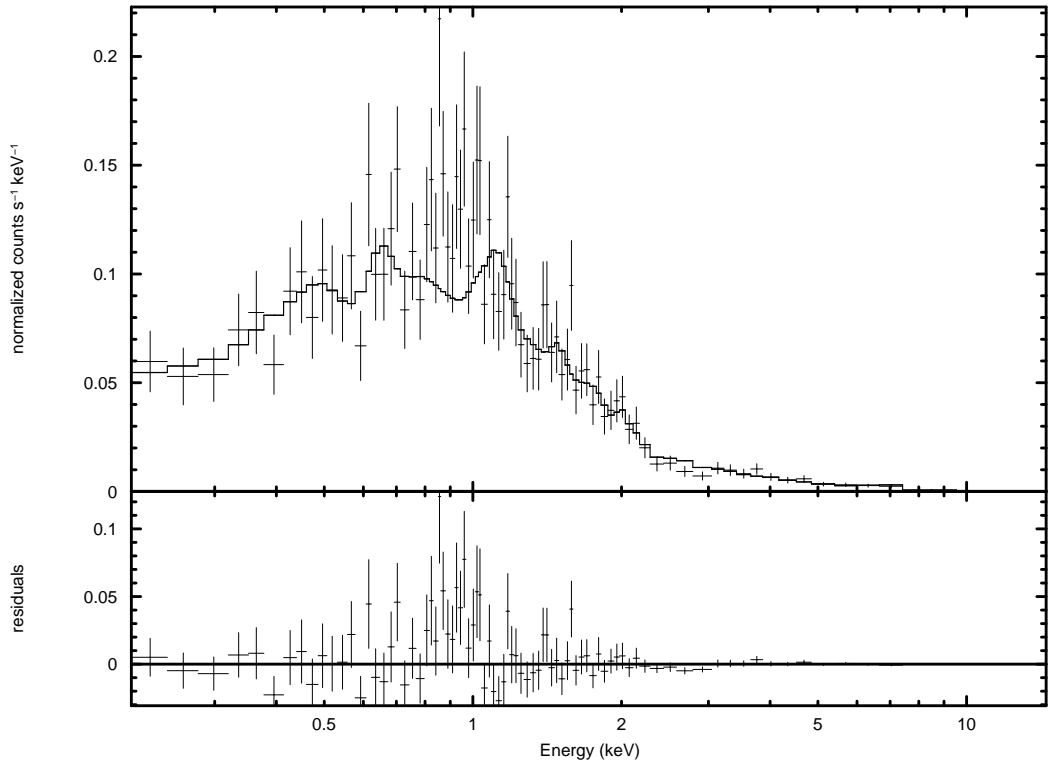}
  \caption{X-ray spectrum of SDSSJ2333 fit with a mekal model with
kT=5 keV.}
\end{figure}

\clearpage
\begin{figure}
  \epsscale{.80}
  \includegraphics[angle=270,scale=.60]{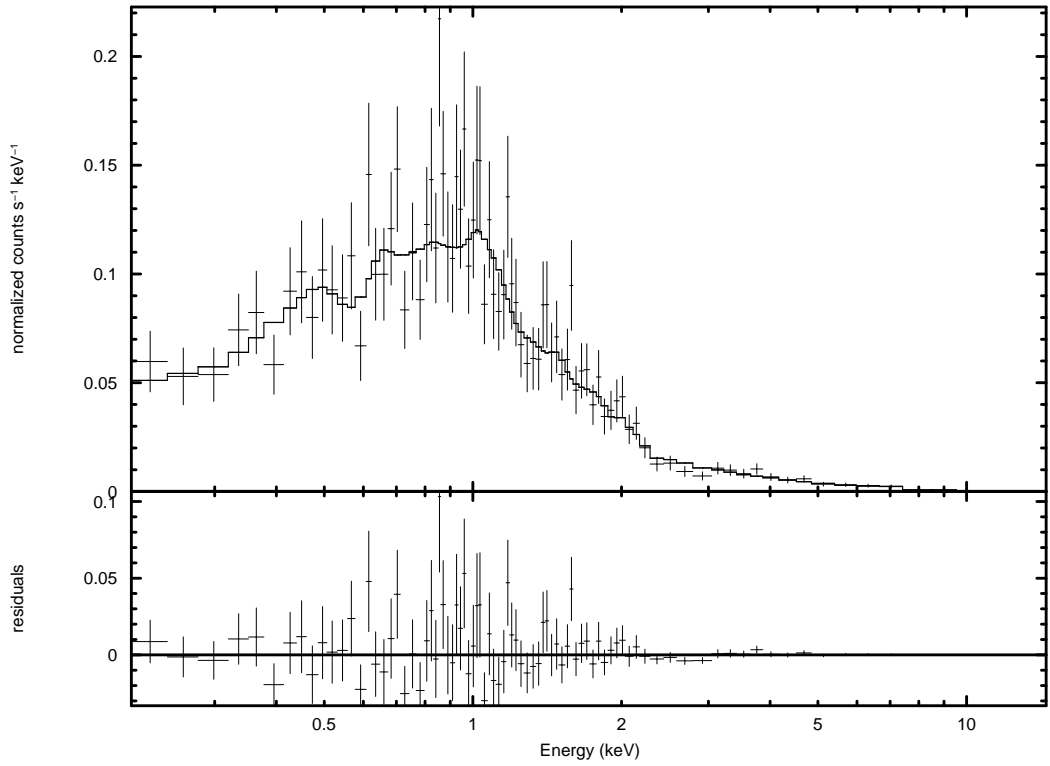}
  \caption{X-ray spectrum of SDSSJ2333 fit with a mkcflow model with
kT from 0.4-17 keV.}
\end{figure}

\clearpage
\begin{figure}
\plotone{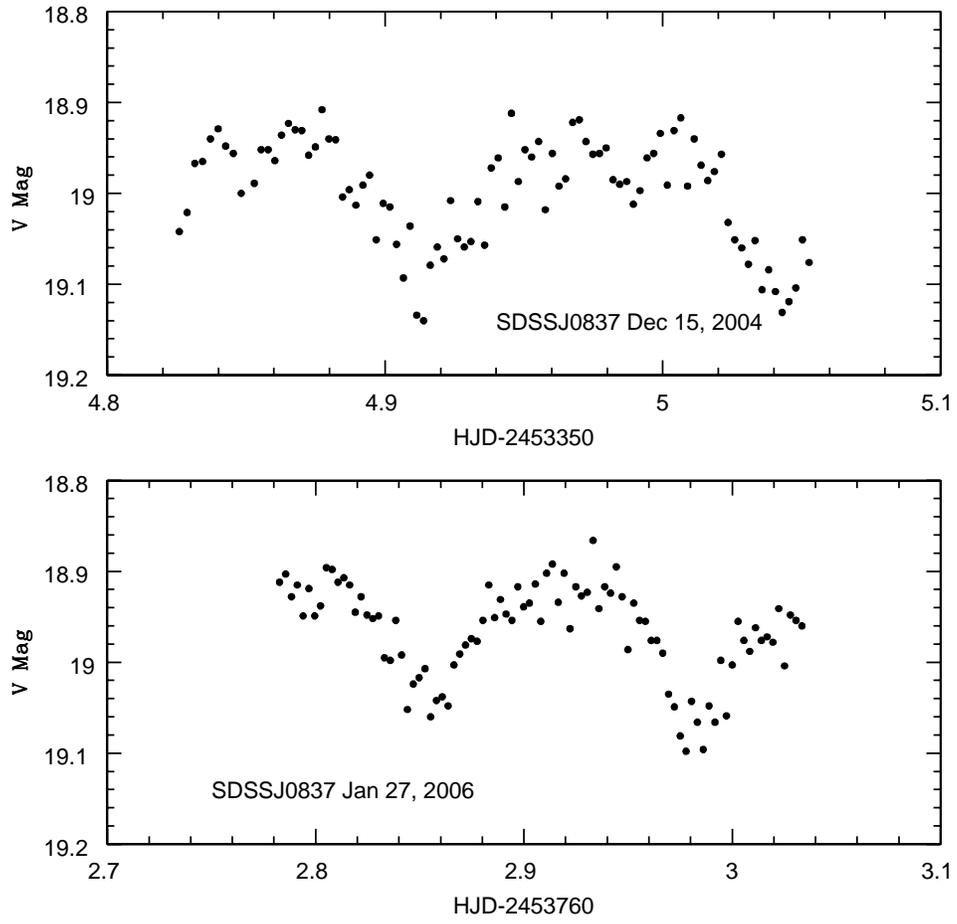}
\caption{Ground-based data on SDSSJ0837. Error bars are 0.03 mag on Dec 15 and
0.02 mag on Jan 27.}
\end{figure}

\clearpage
\begin{figure}
\plotone{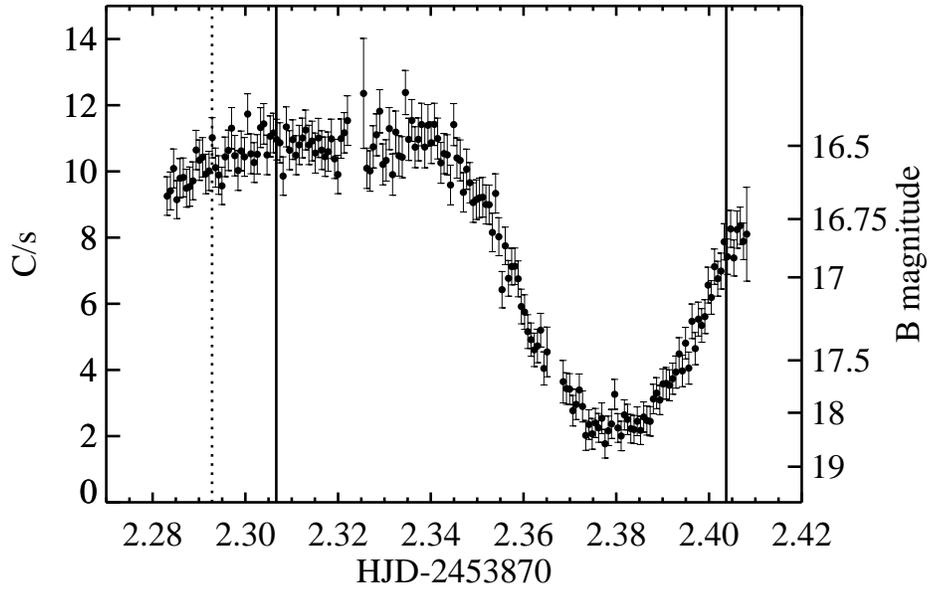}
\caption{OM data on SDSSJ0932 in the B filter. Data are binned into 60 sec
intervals. Dashed line is when the pn observation started; solid lines are
the GTI intervals used for pn.}
\end{figure}

\clearpage
\begin{figure}
\plotone{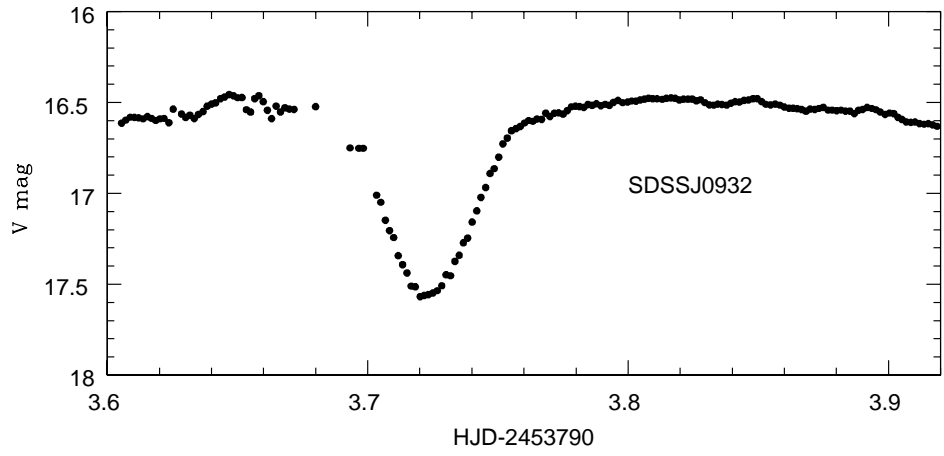}
\caption{Ground-based data on SDSSJ0932 obtained 2006 February 27 UT. Error bars
are 0.01 mag.}
\end{figure}

\clearpage
\begin{figure}
 \epsscale{0.80}
\plotone{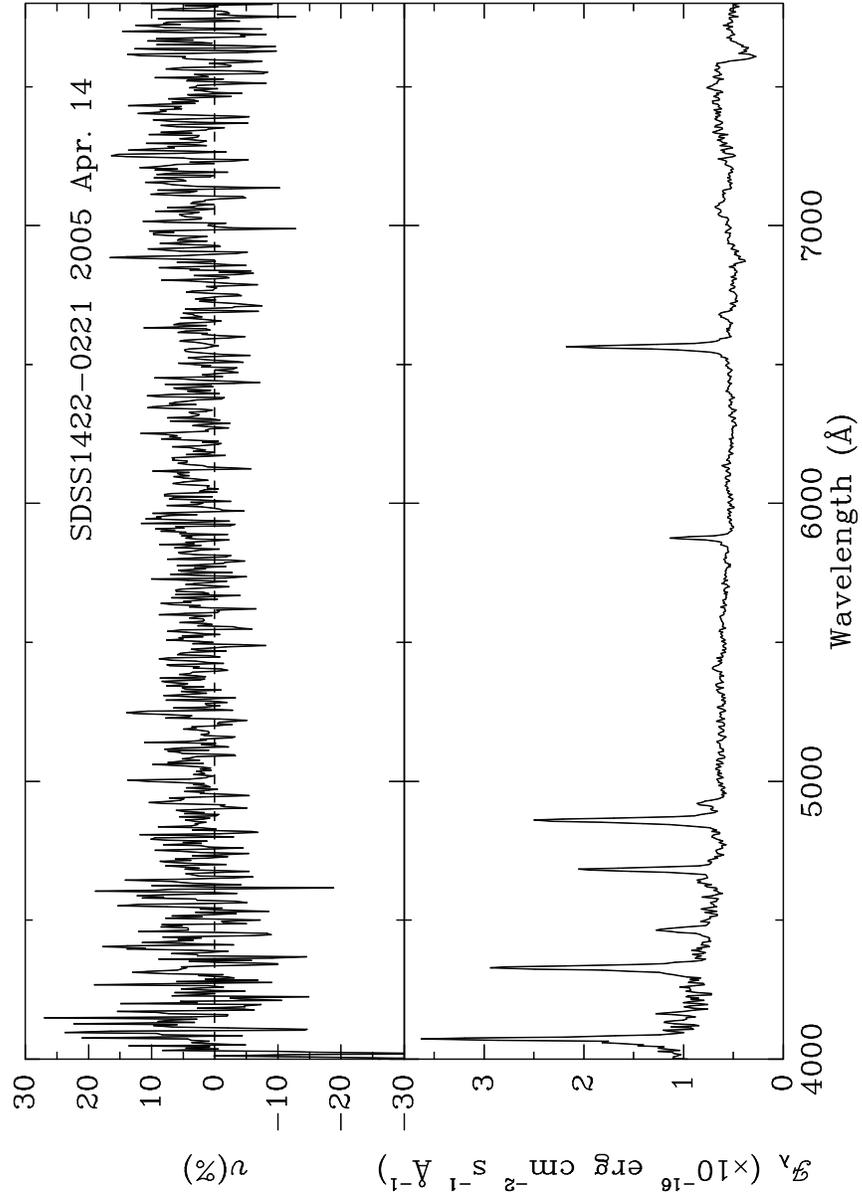}
\caption{Circular polarization (top) and flux (bottom) spectrum of
SDSSJ1422.}
\end{figure}

\begin{figure}
\plotone{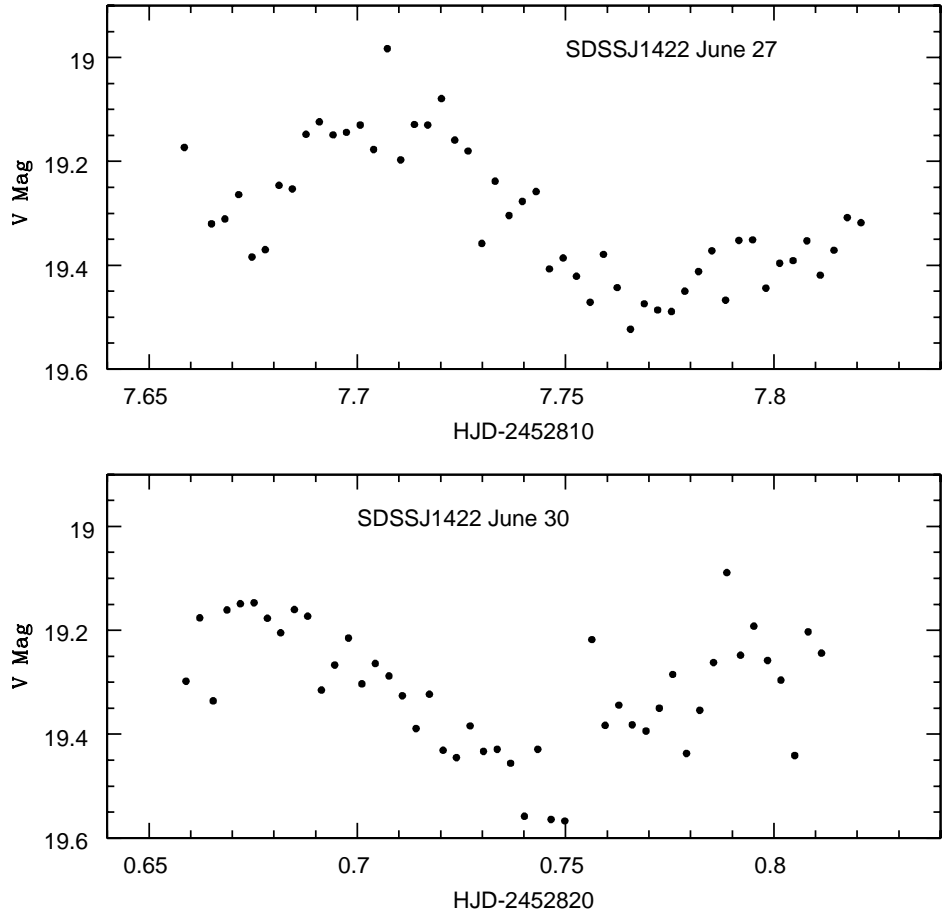}
\caption{Ground-based data on SDSSJ1422. Error bars
are 0.03-0.06 mag.}
\end{figure}

\begin{figure}
\plotone{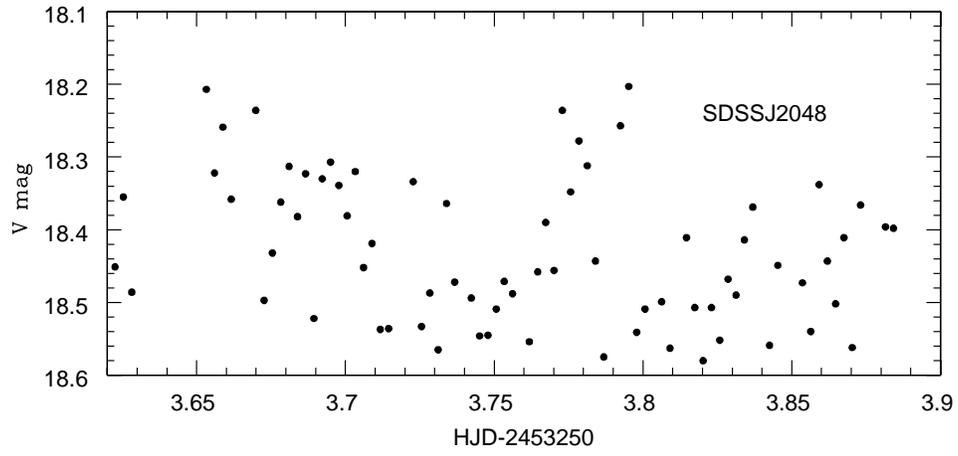}
\caption{Ground-based data on SDSSJ2048 obtained in cirrus. Error bars
are 0.05-0.14 mag.}
\end{figure}

\end{document}